# Système de Communications Sans Fil Très Haut Débit à 60 GHz

Lahatra Rakotondrainibe, Yvan Kokar, Gheorghe Zaharia, Ghais El Zein

IETR-INSA, UMR CNRS 6164
20, Av. des Buttes de Coësmes, CS 14315, 35043 Rennes Cedex, France

lrakoton@insa-rennes.fr; yvan.kokar2@insa-rennes.fr; gheorghe.zaharia@insa-rennes.fr; ghais.el-zein@insa-rennes.fr

**Résumé -** Cet article présente l'étude et la réalisation à l'IETR d'un système de communications sans fil très haut débit à 60 GHz. Le système utilise une architecture monoporteuse simple. La structure du récepteur proposé est basée sur une démodulation différentielle permettant de réduire l'effet des interférences inter-symboles (IIS) et sur une unité de traitement du signal composée d'un bloc de synchronisation conjointe trame et octet et d'un décodeur correcteur d'erreurs RS (255, 239). La technique de synchronisation utilisée permet d'obtenir une forte probabilité de détection de préambule et une très faible probabilité de fausse alarme. Les premiers résultats de mesures ont montré une bonne qualité de transmission pour une liaison en visibilité directe avec des antennes directives.

**Abstract -** This paper presents the study and the realization at IETR of a high data rate 60 GHz wireless communications system. The system uses a simple single carrier architecture. The receiver architecture is based on a differential demodulation which minimizes the intersymbol interference (ISI) effect and a signal processing unit composed of a joint frame and byte synchronization block and a conventional RS (255, 239) decoder. The byte synchronization technique provides a high preamble detection probability and a very small value of the false detection probability. First measurement results show a good communication link quality in line of sight environments with directional antennas.

## 1 Introduction

La montée en fréquence, en particulier à 60 GHz, constitue une des solutions les plus prometteuses pour accroître le débit (quelques Gbit/s) des futurs réseaux locaux sans fil (WPAN). Des études montrant les applications à 60 GHz, la modélisation des canaux de propagation à 60 GHz et les antennes sont citées dans [1-2]. Récemment, les groupes de normalisation IEEE 802.15.3c, ECMA et VHT ont été formés pour normaliser les WPAN dans la bande de 60 GHz [3]. Différentes architectures ont été analysées afin de développer de nouveaux systèmes de communications millimétriques pour des applications multimédia.

Cette étude s'inscrit dans le cadre du projet « Techim@ges ». Ce projet exploratoire s'intéresse à l'étude et la réalisation de systèmes de radiocommunication très haut débit (autour de 1 Gbit/s) à 60 GHz pour de futurs réseaux domestiques. La première application visée est le téléchargement rapide de fichiers volumineux. Le système doit fonctionner en liaison directe dans un environnement domestique. A 60 GHz, l'atténuation provoquée par la traversée d'obstacles tels que les murs est très forte. Compte tenu que la liaison radio à 60 GHz est mono-pièce, une liaison par fibre optique est utilisée pour couvrir toutes les pièces d'un environnement domestique.

Les défauts liés aux blocs radiofréquence (RF) du système (imperfection des composants, bruit de phase, non-linéarité, etc.) peuvent engendrer une dégradation importante des performances. L'intérieur des bâtiments constitue aussi un canal à trajets multiples. Les nombreux obstacles présents (murs, cloisons, plafonds, mobilier) sont autant de surfaces plus ou moins réfléchissantes pour les ondes. L'existence de trajets multiples est la cause du phénomène d'évanouissements et des IIS nécessitant un traitement de signal performant en réception (synchronisation, égalisation, correction d'erreurs), surtout quand le débit est important.

Le suite de l'article contient 3 parties. La partie 2 décrit l'architecture générale du système. La bande de base et le traitement du signal associé sont présentés dans la partie 3. Enfin, la partie 4 traite des résultats de mesure.

## 2 Architecture générale du système

La figure 1 présente le schéma bloc de l'émetteur.

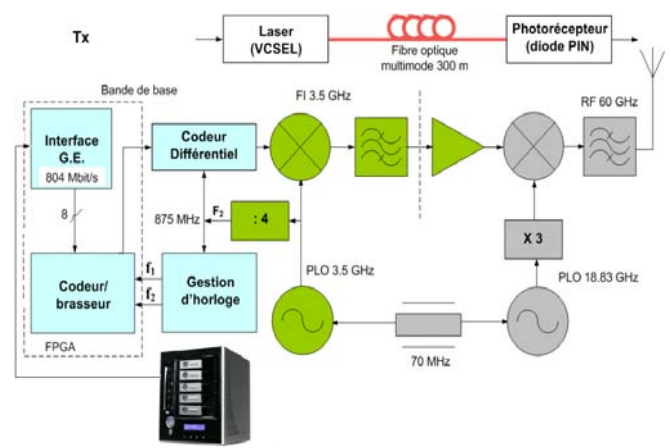

**Figure 1 : Architecture de l'émetteur (Tx) à 60 GHz**

La modulation MDP2-D est simple et permet de résoudre le problème d'ambiguïté de phase en réception grâce à une démodulation différentielle. Comparée à l'OFDM, cette technique est moins sensible à l'instabilité et au bruit de phase de l'oscillateur local

(OL), ainsi qu'aux non-linéarités de l'amplificateur de puissance.

En émission, après les opérations de codage de canal et de brassage, le flux d'information est appliqué à l'entrée du codeur différentiel. Les symboles codés vont ensuite moduler une porteuse en FI à 3.5 GHz générée par un oscillateur verrouillé en phase (PLO). Le signal modulé en FI est filtré dans une bande de 2 GHz puis transmis par une fibre optique d'environ 300 m. Ce signal FI est utilisé pour moduler directement le courant d'un laser émettant à 850 nm. Le signal optique est reconverti en signal électrique au moyen d'une diode PIN. A la sortie de la fibre optique, le signal est amplifié et transmis au bloc RF d'émission à 60 GHz. Ce bloc est composé d'un mélangeur, d'un tripleur de fréquence, d'un PLO à 18.83 GHz et d'un filtre passe-bande (59-61 GHz). Le bruit de phase du PLO à 18.83 GHz est de - 110 dBc/Hz à 10 kHz de la porteuse. Le signal RF de puissance 0 dBm est transmis via une antenne cornet avec un gain de 22.4 dBi et un angle d'ouverture de 12°.

La figure 2 présente le schéma bloc du récepteur.

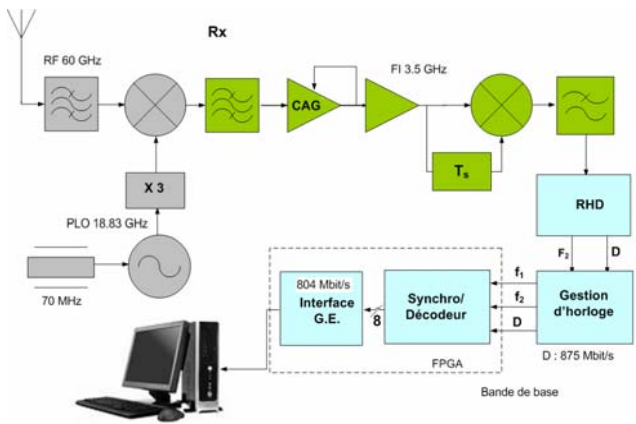

**Figure 2 : Architecture du récepteur (Rx) à 60 GHz**

En réception, le signal RF est d'abord filtré pour limiter l'effet du bruit à l'entrée du récepteur. Le signal obtenu est transposé en FI à 3.5 GHz et filtré. Un contrôle automatique de gain (CAG) ayant une dynamique de 20 dB permet d'assurer un signal avec un niveau quasi-constant à l'entrée du démodulateur. On utilise une démodulation différentielle qui assure à la fois la démodulation et le décodage par transition. Comparée à une démodulation cohérente, la démodulation différentielle est une solution techniquement plus simple, mais moins performante dans un canal à Bruit Blanc Additif Gaussien (BBAG). Par contre, elle s'avère moins sensible à l'influence des trajets multiples et par conséquent aux interférences entre symboles qu'ils peuvent engendrer. Grâce au produit de deux symboles consécutifs, le rapport entre le lobe principal et le deuxième lobe de la réponse impulsionnelle du canal est élevé au carré. Par conséquent, les autres trajets sont fortement atténués par rapport au premier trajet. Un filtrage passe-bas (fréquence de coupure 1 GHz) permet de restituer le signal en bande de base. La récupération d'horloge se fait à partir des données reçues remises en forme. De longues séquences de 0 ou de 1 consécutifs sont à éviter pour permettre le bon fonctionnement du circuit de récupération d'horloge et de données (RHD). Par conséquent, il faut brasser les données en émission et les débrasser en réception.

## 3 Architecture en bande de base (BB)

Les blocs BB-Tx et BB-Rx sont implémentés dans un circuit programmable FPGA Xilinx Virtex 4. Le bloc BB-Tx comporte le codage de canal, la formation de la trame et le brassage des données à transmettre, comme le montre la figure 3.

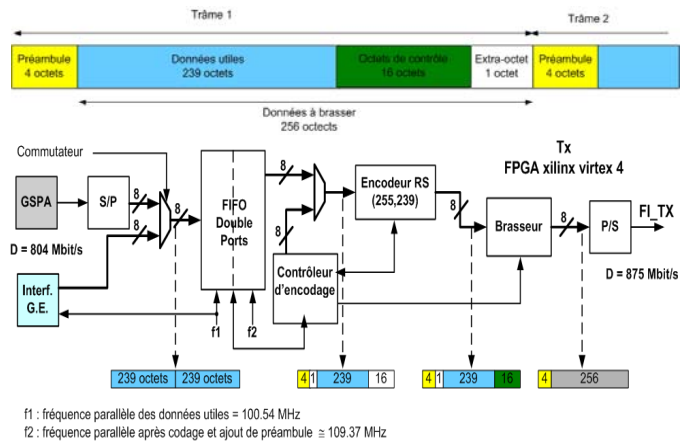

**Figure 3 : Format de la trame et architecture de BB-Tx**

Le codage de canal est réalisé avec un codeur RS(255,239), travaillant en octet. Lorsque la source de données est en série, une conversion série/parallèle (S/P) est nécessaire. Il est possible d'accepter des données par octet fournies par une interface Gigabit Ethernet (G.E.). La trame à transmettre est formée de 4 octets de préambule, 239 octets de données utiles, 16 octets de contrôle et 1 extra-octet afin d'assurer une trame contenant un nombre d'octets (sauf le préambule) multiple de 8 (pour simplifier les opérations de brassage/débrassage). Comme la trame à transmettre après codage est de 260 octets, il est indispensable de disposer de deux fréquences $f_1$ et $f_2$, générées dans le bloc gestion d'horloge, telles que :

$$f_1 = \frac{F_1}{8} = 100.54 \text{ MHz}, \quad f_2 = \frac{F_2}{8} = 109.37 \text{ MHz}.$$

où : $F_2 = \frac{IF}{4} = 875 \text{ MHz}$ et $\frac{F_1}{F_2} = \frac{239}{260}$.

Pour former la trame, les octets sont écrits dans la mémoire FIFO double port avec la fréquence $f_1$. Dès que la mémoire FIFO est à moitié pleine (pour compenser les écarts de fréquence) les données sont lues à la fréquence $f_2$. Le contrôleur de codage génère 4 octets de préambule et vient lire 239 octets (nombre d'octets utiles dans une trame de codage). Le codeur RS lit 1 octet à chaque coup d'horloge. Après 239 périodes d'horloge, le contrôleur de codage arrête le transfert de données pendant 17 coups d'horloge afin d'y ajouter les 16 octets de contrôle propres au codage RS et l'extra-octet. Le brassage des données est réalisé avec une séquence de 8 octets (une séquence pseudo-aléatoire de 63 bits + 1 bit additionnel). Le brasseur reçoit une trame dont les 4 octets du préambule ne seront pas brassés.

Avant le préambule P de 4 octets, on place un extra-octet additionnel noté : d = [d(1) d(2) … d(8)].
La question est de savoir quel devrait être le contenu binaire de cet octet capable d'assurer la plus faible corrélation possible. Ce problème est posé sans considérer les erreurs aléatoires qui peuvent apparaître lors de la transmission des données. On note :

$$k = d(8)*2^7+d(7)*2^6+\ldots+d(2)*2+d(1), \quad 0 \leq k \leq 255.$$

La plus grande valeur de la corrélation entre P et la trame formée par P et d (dans la trame, le préambule P est suivi par l'extra-octet dans l'ordre d(1), d(2), etc.) est représentée par la figure 4 en fonction de k :

$$\text{Mcor}(k) = \max_i \left\{ \text{sum}(\overline{P \oplus T}) \right\}$$

où $P = [P(1)\ P(2)\ldots\ P(32)]$
et $T = [d(9-i)\ldots d(2)\ d(1)\ P(1)\ P(2)\ldots P(32-i)]$, $1 \leq i \leq 8$.

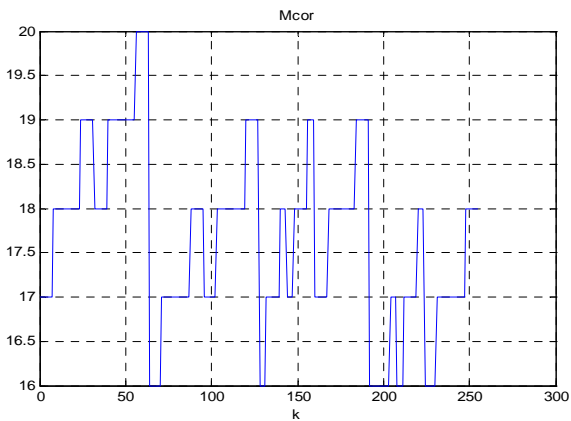

**Figure 4 : Evolution de Mcor (k) en fonction de k**

Pour chaque valeur k on calcule 8 valeurs de cette fonction de corrélation, en fonction du nombre i de bits de l'octet d considérés avant le préambule P amputé de ses derniers i bits.

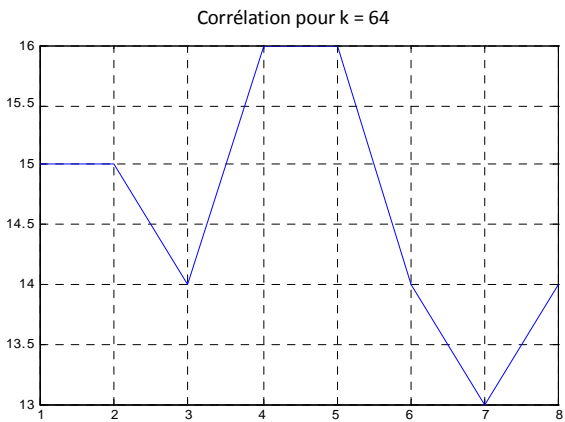

**Figure 5 : Evolution de la fonction de corrélation entre P et T**

La meilleure situation indiquée à la figure 5 est obtenue pour k = 64. On a : d = [0 0 0 0 0 0 1 0], donc seulement d(7) = 1, les autres bits sont tous égaux à 0.
La figure 6 présente l'architecture de BB-Rx. En réception, le flux de données série à 875 Mbit/s en sortie de RHD est parallélisé sur 8 bits. Les procédés de détection du préambule et du contrôle d'alignement sont présentés dans la figure 7.

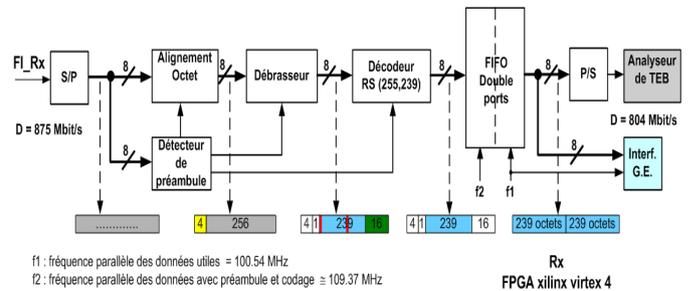

**Figure 6 : Architecture de BB-Tx**

La méthode consiste à effectuer une corrélation entre chaque séquence de 32 bits successifs reçus et le préambule (4 octets). On décale ensuite d'un bit pour chaque corrélateur la séquence de données à analyser. En tout, pour la détection d'un préambule et la synchronisation octet, il y a 8 corrélateurs de 32 bits. Par conséquent, on effectue la détection d'un préambule à l'aide de 32 + 7 = 39 bits (+ 7 à cause des différents décalages possibles d'un octet). Pour obtenir une faible probabilité de non-detection de préambule et de fausse alarme [4], on tient compte de la périodicité des préambules. En détectant 2 préambules successifs, chaque valeur de la fonction de corrélation entre le préambule et la séquence des données reçues est comparée à un seuil à déterminer.

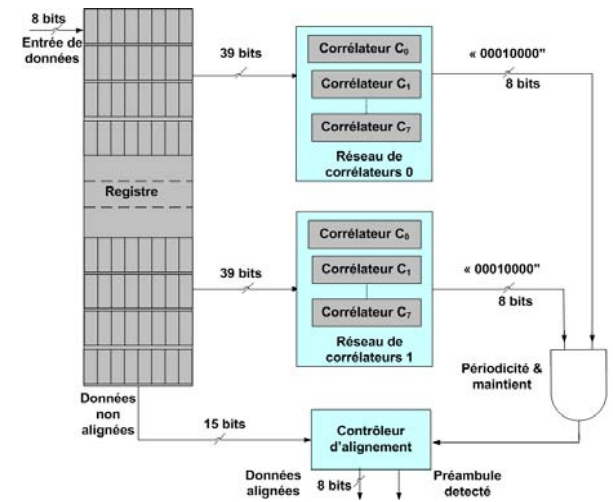

**Figure 7 : Détection du préambule et du contrôle d'alignement**

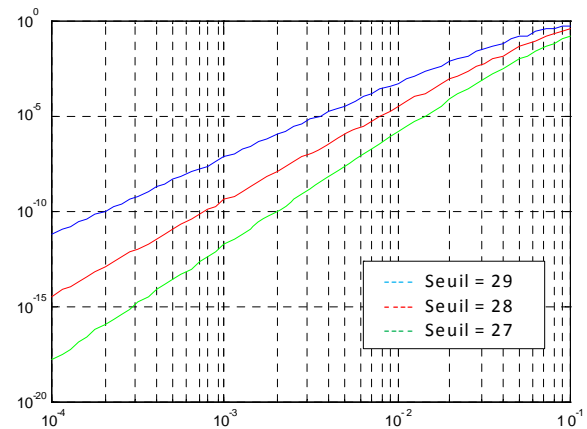

**Figure 8 : Probabilité de détection du préambule en fonction de la probabilité d'erreurs dans le canal**

Une fausse alarme est déclarée lorsque chacun des 2 corrélateurs placés à la bonne distance (celle entre P3 et

P2 ou celle entre P2 et P1, ce qui est la même chose, compte tenu de la périodicité des préambules) indique la détection de préambule dans les 256 octets de données. L'analyse commence avec la séquence [P3 D2 d2 P2] décalée d'un bit vers la droite (afin d'éviter une vraie détection du préambule) et se termine avec [P2 D1 d1 P1] décalée d'un bit vers la gauche. Le nombre d'apparitions des fausses alarmes dépend du seuil S qui prend des valeurs de 1 à 32, selon la figure 9.

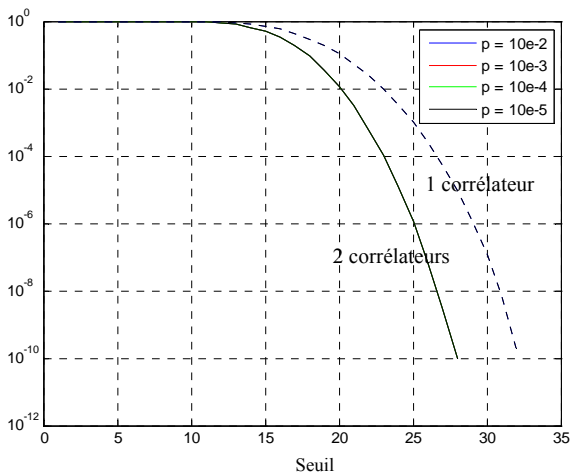

**Figure 9 : Probabilité de fausse alarme**

La prise de décision doit se faire à partir de 264 octets successifs (préambule + données + préambule) stockés dans un registre. Dès que les corrélateurs de même rang $C_k$ des deux réseaux de corrélateurs indiquent la détection de préambule, l'opération est validée. Par conséquent, à partir de $8 + 7 = 15$ bits successifs, on peut assurer l'alignement octet.

Dès que l'alignement octet est assuré et le débrassage effectué, le décodeur RS calcule pour chaque trame reçue une pluralité de syndromes d'erreurs représentatifs du nombre et des emplacements des erreurs détectées dans cette trame. En fonction de ces syndromes, le décodeur détermine les corrections à effectuer, dans la mesure où le nombre d'erreurs ne dépasse pas la capacité de correction du code (8 octets). Finalement, les données sont fournies sur 8 bits à l'interface G.E.

## 4 Résultats de mesures

Afin de valider le système, une transmission de données à 875 Mbit/s a été réalisée en visibilité directe. Selon la figure 10, le diagramme de l'œil obtenu montre une bonne ouverture (pour une distance Tx-Rx de 10 m) traduisant une bonne qualité de transmission. Afin de déterminer la qualité de la liaison, des mesures de TEB ont été effectuées avec deux types d'antennes : antennes cornet décrites précédemment et antennes imprimées avec un gain de 8 dBi et un angle d'ouverture de 30°. Les résultats de TEB sont donnés par la figure 11. Ces résultats de mesure montrent qu'avec des antennes cornets, les performances du système sont améliorées par rapport à des antennes imprimées. Cependant, s'il y a un obstacle, la communication peut être coupée.

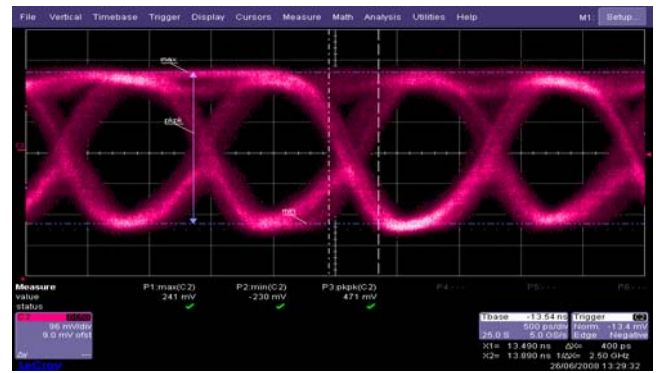

**Figure 10 : Diagramme de l'œil à 875 Mbit/s (Tx-Rx : 10 m)**

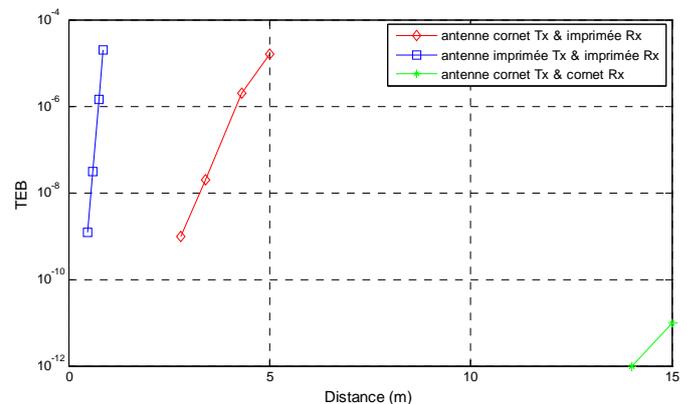

**Figure 11: Résultats de mesure de TEB (sans codage)**

## 5 Conclusions

Ce papier a présenté la conception et la réalisation d'un système de communications sans fil très haut débit à 60 GHz. Les premiers résultats de mesure ont montré une bonne qualité de transmission en liaison visibilité directe avec des antennes directives. Avec des antennes moins directives, l'ajout d'une égalisation est essentiel pour combattre la sélectivité fréquentielle du canal. De même, il est prévu de mettre en œuvre des modulations à plus grand nombre d'états pour améliorer l'efficacité de transmission.

### Références

[1] P. Smulders, H. Yang, I. Akkermans, "On the design of Low-Cost 60 GHz Radios for Multigigabit-per-Second Transmission over Short Distances", *IEEE Communications Magazine*, 8 pages, déc. 2007.

[2] S. Collonge, G. Zaharia, and G. El Zein, "Wideband and Dynamic Characterization of the 60 GHz Indoor Radio Propagation-future Home WLAN Architectures", *Annals of Telecommunications*, Vol. 58, N° 3-4, pp. 417-447, mars-avril 2003

[3] R. Funada et al., "A design of single carrier based PHY for IEEE 802.15.3c standard", *Proc. of the IEEE PIMRC 2007*, 3-7 sept 2007

[4] H. Urkowitz, *Signal Theory and Random Processes*, Artech House, 1983, pp. 505-542